\documentclass[aps,showpacs,prb,reprint,superscriptaddress,longbibliography]{revtex4-1}

\usepackage[colorlinks=true,linkcolor=blue,citecolor=blue,urlcolor=blue]{hyperref}
\usepackage{setspace} 
\usepackage{graphicx}
\usepackage{amsmath}
\usepackage{color}
\usepackage{amssymb}
\usepackage{verbatim}
\usepackage{latexsym}
\usepackage{enumerate} 
\usepackage{bm} 
\usepackage{caption}
\usepackage{subcaption}
\usepackage{mathtools}
\usepackage{braket}
\usepackage{multirow}

\begin{document}

\title{First-principles anharmonic vibrational study of the structure of calcium silicate perovskite under lower mantle conditions}

\author{Joseph C.\ A.\ Prentice}
\affiliation{Departments of Materials and Physics, and The Thomas Young
  Centre for Theory and Simulation of Materials, Imperial College
  London, London SW7 2AZ, United Kingdom}
\affiliation{TCM Group, Cavendish Laboratory, University of Cambridge,
  J.\ J.\ Thomson Avenue, Cambridge CB3 0HE, United Kingdom}
\author{Ryo Maezono}
\affiliation{School of Information Science, JAIST, 1-1 Asahidai, Nomi, Ishikawa 923-1292, Japan}
\author{R.\ J.\ Needs}
\affiliation{TCM Group, Cavendish Laboratory, University of Cambridge,
  J.\ J.\ Thomson Avenue, Cambridge CB3 0HE, United Kingdom}

\date{\today}

\begin{abstract}

Calcium silicate perovskite (CaSiO$_3$) is one of the major mineral
components of the lower mantle, but has been the subject of relatively
little work compared to the more abundant Mg-based materials. One of
the major problems related to CaSiO$_3$ that is still the subject of
research is its crystal structure under lower mantle conditions -- a
cubic \emph{Pm$\bar{3}$m} structure is accepted in general, but some
have suggested that lower-symmetry structures may be relevant. In this
work, we use a fully first-principles vibrational self-consistent
field (VSCF) method to perform high accuracy anharmonic vibrational
calculations on several candidate structures at a variety of points
along the geotherm near the base of the lower mantle, in order to
investigate the stability of the cubic structure and related distorted
structures. Our results show that the cubic structure is the most
stable throughout the lower mantle, and that this result is robust
against the effects of thermal expansion.
  
\end{abstract}

\maketitle

\section{Introduction}

Of the various layers that make up the internal structure of the
Earth, the lower mantle is by far the largest by volume, constituting
around 55\% of the volume of the entire
Earth\cite{dziewonski_preliminary_1981}. Its upper boundary is marked
by the `transition zone' between the upper and lower mantle at depths
of $410$-$660$~km, whilst its lower edge lies at the core-mantle
boundary (CMB) at a depth of
$2890$~km.\cite{jordan_structural_1979,burns_mineralogical_1993} The
composition of the lower mantle and the structures of the minerals
within it is still subject to
debate\cite{mattern_lower_2005,bovolo_physical_2005}.

The main elements present in the mantle as a whole are magnesium,
iron, calcium, silicon, aluminium and oxygen, with the upper mantle,
transition zone, and lower mantle differentiating themselves through
the minerals that these elements form. Olivine ((Mg,Fe)$_2$SiO$_4$),
which dominates the upper mantle, undergoes a series of phase
transitions within the transition zone, ending with the minerals
Mg-perovskite and magnesiow\"{u}stite, which have chemical formulae
(Mg,Fe)SiO$_3$ and (Mg,Fe)O respectively. These two minerals
constitute over 80\% of the lower mantle between
them\cite{irifune_absence_1994}. More recent work has also identified
a further $200$~km thick layer directly above the CMB, known as the
D$''$ layer, which exhibits more complex behaviour. The D$''$ layer
has been the subject of considerable study, and is thought to arise
from a further phase transition, from Mg-perovskite to the
`post-perovskite'
phase\cite{oganov_theoretical_2004,murakami_post-perovskite_2004,tsuchiya_phase_2004}. This
is a layered CaIrO$_3$-type structure with orthorhombic \textit{Cmcm}
symmetry, whereas the perovskite structure has cubic
\textit{Pm$\bar{3}$m} symmetry.

Beyond Mg-perovskite and magnesiow\"{u}stite, the third most
significant mineral by volume in the lower mantle is calcium silicate
perovskite,
CaSiO$_3$\cite{irifune_absence_1994,caracas_theoretical_2006}. Much
work has been focused on the magnesium and iron-based minerals, and
the behaviour of this less common but still significant material has
been relatively neglected, although it is thought to have an effect on
the shear velocities of seismic waves as they travel through the
Earth, potentially having significance for understanding
earthquakes\cite{greaux_sound_2019,caracas_casio3_2005,mattern_lower_2005,karki_first-principles_1998}. However,
the most basic property of calcium silicate -- its crystal structure
-- is still the subject of research efforts. Understanding the
structure of calcium silicate in the lower mantle is vital, as
changing the structure may have significant effects on the seismic
velocity in the
material\cite{karki_first-principles_1998,uchida_non-cubic_2009}. In
particular, it has been shown that a tetragonal distorted structure of
CaSiO$_3$ has a significantly lower S-wave speed than the cubic
structure\cite{stixrude_phase_2007,kudo_sound_2012,greaux_sound_2019}.

The fundamental structure of any perovskite material is cubic, and
contains a standard motif of corner-sharing oxygen octahedra.  In many
perovskite-class materials, however, this perfect cubic structure may
be unstable with respect to various symmetry-breaking distortions,
depending on the identities of the ions involved. These instabilities
reveal themselves as soft modes of the cubic stucture, often involving
partial rotations of the octahedra. Previous work on the existence of
such instabilities in calcium silicate has produced mixed results,
both from a theoretical and an experimental perspective. Although it
was thought initially that the cubic structure was the ground state
\cite{wentzcovitch_ab_1995,hemley_theoretical_1987,liu_synthesis_1975,wang_thermal_1996},
subsequent theoretical and experimental work demonstrated the
existence of soft modes in the cubic
structure\cite{stixrude_prediction_1996,caracas_theoretical_2006,stixrude_phase_2007,jung_ab_2005},
leading to experimental demonstrations that the ground state is a
distorted cubic
structure\cite{shim_tetragonal_2002,kurashina_phase_2004,ono_phase_2004}. Unfortunately,
however, the distortions are small enough that there is a degree of
experimental uncertainty in the structure of the lowest energy
distorted state, with both
tetragonal\cite{shim_tetragonal_2002,stixrude_phase_2007,komabayashi_phase_2007,caracas_theoretical_2006}
and
orthorhombic\cite{uchida_non-cubic_2009,magyari-kope_low-temperature_2002,adams_ab_2006}
phases proposed. Distorted structures with several different
symmetries have been studied, including those with space groups
\emph{I4/mcm}, \emph{Imma}, \emph{P4/mbm}, \emph{I4/mmm}, \emph{Im3},
\emph{P4$_2$/nmc} and \emph{Pnma}\cite{caracas_theoretical_2006}. It
is believed that the cubic \emph{Pm$\bar{3}$m} phase will be
stabilised at some temperature and pressure, but different groups have
obtained different results for these transition conditions -- although
it is generally accepted that calcium silicate is cubic in most of the
lower
mantle\cite{nestola_casio3_2018,sun_dynamic_2014,komabayashi_phase_2007,stixrude_phase_2007,adams_ab_2006,noguchi_high-temperature_2012},
there have been suggestions that distorted structures are also
relevant in some
regions\cite{uchida_non-cubic_2009,ono_phase_2004,li_phase_2006}.
Intriguingly, it has even been suggested that a phase transition from
a cubic structure \textit{into} a distorted structure may occur at
very high pressures, seen at the very base of the
mantle\cite{hemley_theoretical_1987}. Previous studies have suggested
that anharmonic vibrational effects are significant in this
system\cite{sun_dynamic_2014,hemley_theoretical_1987}, implying that
high accuracy anharmonic calculations will be necessary to discover
the correct structure of calcium silicate in the lower mantle. In the
present work, we concentrate on the bottom of the lower mantle for
several reasons: the presence of the D$''$ layer, the previous
suggestion of a cubic to distorted phase transition, and the
difficulty of performing experimental measurements at relevant
conditions, meaning first-principles calculations have a vital role to
play.

\begin{figure}
\centering
\includegraphics[trim={6cm 1cm 6cm 0cm},clip,width=0.45\textwidth]{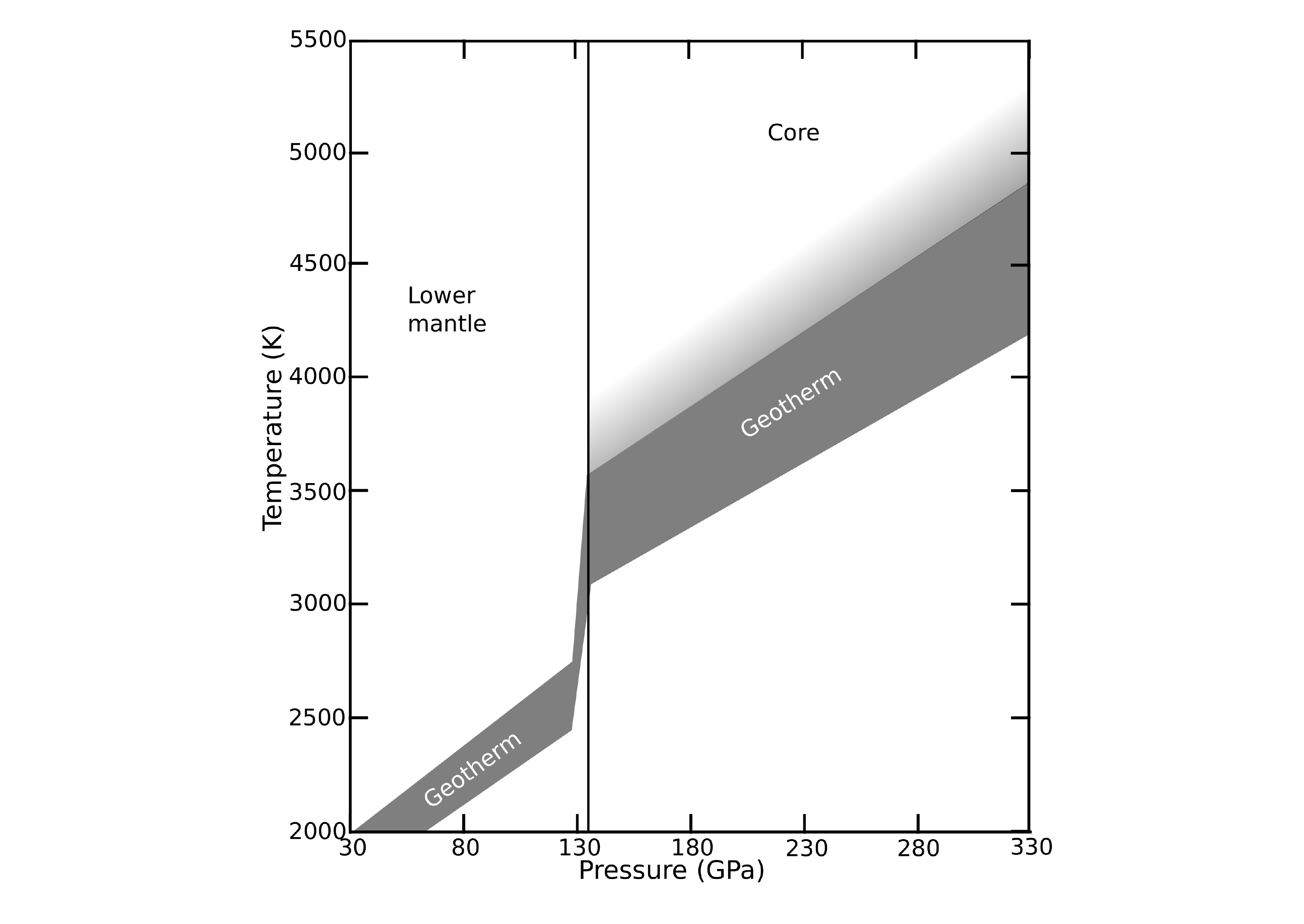}
\caption{The currently accepted form of the geotherm in the lower
  mantle and outer core, showing the conditions that can be found in
  this region of the Earth's interior. Note the significant
  discontinuity in the gradient of the geotherm near the core-mantle
  boundary (CMB), marked by the line at around $136$~GPa, and the
  steeper section at pressures just below the CMB corresponding to the
  D$''$ layer. Figure adapted from
  Ref.\ \onlinecite{nomura_low_2014}.}
\label{fig:Geotherm}
\end{figure}

The conditions within the layers in the Earth vary significantly with
depth. In particular, with increasing depth, the pressure and
temperature increase, although the relationship between the two is not
necessarily simple. The geotherm, shown in Fig.\ \ref{fig:Geotherm},
describes the relationship between temperature and pressure (or depth)
within the Earth\cite{nomura_low_2014}. A given depth corresponds to a
set of temperature and pressure ($T,p$) conditions on the geotherm. To
obtain the correct structure of a mineral at this depth, the free
energy of the system must be minimised under this set of
conditions. Experimentally, it is often difficult to control both
temperature and pressure simultaneously if conditions deep within the
Earth are of interest, making accurate first-principles calculations
extremely valuable for this purpose.

In this work, we go beyond previous work on this system by approaching
it from a fully \textit{ab initio} standpoint -- we use highly
accurate first-principles density functional theory (DFT)
calculations, together with a vibrational self-consistent field method
for anharmonic vibrational calculations.  Using these methods, we
examine the relative stability of three possible crystal structures of
calcium silicate, including the cubic perovskite structure, at four
points on the geotherm, including the effects of anharmonic
vibrations. This investigation constitutes the most in-depth and
accurate investigation of this important topic yet attempted. Our
results suggest that, although the cubic phase of calcium silicate is
unstable to distortions at zero temperature, calcium silicate takes up
the cubic structure throughout the lower part of the lower mantle. The
effects of thermal expansion are also considered, and found not to
affect the conclusions of this work.

The rest of this work is organised as follows: in
Sec.~\ref{sec:Methods} we outline the computational methods employed
in this work, and give technical details of the calculations. In
Sec.~\ref{sec:Results} we present the main results of our study,
detailing the structures considered at different points along the
geotherm, their relative stabilities, and the effect of anharmonicity
on these results. In Sec.~\ref{sec:Summary} we give a brief summary of
our results and some concluding remarks.

\section{Calculational methods} \label{sec:Methods}

In this work, as we are studying systems along the geotherm, the
effect of both temperature and pressure must be included when
comparing different structures. This means that the appropriate
quantities to compare are the enthalpy $H$ rather than the energy $E$
at zero temperature, and the Gibbs free energy $G$ rather than the
Helmholtz free energy $F$ at finite temperature, where $H=E+pV$ and
$G=F+pV$.

\subsection{DFT calculations}

All DFT calculations in this work were performed using version 16.1 of
the {\sc castep} code\cite{clark_first_2005}, with the corresponding
`on-the-fly' ultrasoft pseudopotentials\cite{vanderbilt_soft_1990} and
the PBE exchange-correlation
functional\cite{perdew_generalized_1996}. The PBE functional was
chosen as it has been used in previous calculations under lower mantle
conditions\cite{jung_ab_2005,oganov_theoretical_2004,pickard_structures_2015}. A
cut-off energy of $1000$~eV was used in all calculations, and
Monkhorst-Pack grids\cite{monkhorst_special_1976} with spacings as
close to $0.025$~$\text{\r{A}}^{-1}$ as possible were used throughout
-- this corresponded to a $13\times13\times13$ grid in the unit cell
of all the undistorted cubic structures considered. The total energy
was always converged to within $10^{-9}$~eV, to ensure accurate forces
for the vibrational calculations. Within geometry optimisation
calculations, the atomic positions were adjusted until the root mean
square of the forces on all the atoms was below $0.0001$~eV/\r{A}.

\subsection{Vibrational calculations}

In this work a vibrational self-consistent field (VSCF) method,
described in Ref.~\onlinecite{monserrat_anharmonic_2013} and used
successfully several times
since\cite{azadi_dissociation_2014,monserrat_electron-phonon_2014,engel_anharmonic_2015,prentice_first-principles_2017},
is used to include anharmonic effects in our results. This method has
been used successfully in high pressure and temperature
systems\cite{monserrat_electron-phonon_2014,monserrat_structure_2018},
but this study consitutes the first time such a method has been
applied to lower mantle materials. The method uses a basis of harmonic
normal mode co-ordinates to describe the Born-Oppenheimer (BO) surface
that the nuclei move in, but does not enforce the harmonic
approximation itself. Instead, we go beyond the harmonic
representation of the BO surface by using a principal axes
approximation,\cite{jung_vibrational_1996} resulting in a many-body
term expansion of the BO surface:
\begin{equation} 
  E(\mathbf{u})=E(\mathbf{0})+\sum_i V_1(u_i) + \frac{1}{2} \sum_{i'\neq i} V_2(u_i,u_{i'}) + \cdots \label{eq:BOSurfExp}
\end{equation}
Here, $\mathbf{u}$ is a collective vector containing the normal mode
amplitudes $u_i$, Latin indices are collective labels for the quantum
numbers $(\mathbf{q},\nu)$ where $\mathbf{q}$ is a phonon wavevector
and $\nu$ a phonon branch, and $E(\mathbf{u})$ is the energy of the BO
surface when the atomic nuclei are in configuration
$\mathbf{u}$. Anharmonicity is already included in the $V_1$ terms, as
they are not constrained to the harmonic form. In this work, this
expansion is truncated to only include the $V_1$ terms, as corrections
due to higher order terms in this expansion have been found to be less
important\cite{monserrat_anharmonic_2013,prentice_using_2017}. The
various $V_1$ terms are found by mapping the BO surface as a function
of the amplitude of each normal mode, typically using DFT calculations
with the normal mode frozen in, and then fitting a spline to the
results. This provides a truly first-principles description of the BO
surface, without assuming a functional form, and provides significant
amounts of information about the surface the nuclei move in. Mapping
the BO surface is by far the most computationally expensive part of
the VSCF method, as a large number of DFT calculations are required to
achieve an accurate fit. Once a form for the BO surface is obtained,
the resulting nuclear Schr\"{o}dinger equation is solved numerically
(and self-consistently) to obtain the anharmonic energies and
wavefunctions for both ground and excited vibrational states. These
can then be used to calculate the anharmonic vibrational free energy.

In order to obtain the basis of harmonic normal modes required for the
VSCF method, a harmonic vibrational calculation was performed. This
used a finite displacement method to calculate the matrix of force
constants, which was then Fourier transformed to obtain the dynamical
matrix.\cite{kunc_ab_1982} This was diagonalised to obtain the
harmonic vibrational frequencies and eigenvectors. Atomic
displacements of $0.00529$~\r{A} were used. To reduce the
computational cost of the mapping of the BO surface within the VSCF
method itself, the non-diagonal supercells method was used to reduce
the size of the supercell needed to sample a given point in the
vibrational Brillouin zone\cite{lloyd-williams_lattice_2015}, and DFT
force data was used to give a more accurate fit for a given number of
mapping calculations\cite{prentice_using_2017}. 21 different
amplitudes per mode were used to map the BO surface. For the purpose
of self-consistently solving the nuclear Schr\"{o}dinger equation, the
vibrational wavefunction $\ket{\Phi(\mathbf{u})}$ is written as a
Hartree product of the normal modes, $\prod_i \ket{\phi_i (u_i)}$.

The states $\ket{\phi_i (u_i)}$ are represented in a basis of
one-dimensional harmonic oscillator eigenstates. To be able to get
accurate results for the extreme temperatures considered in this work,
it is necessary to have a good description of the high energy
vibrational states and their vibrational wavefunctions, which itself
requires a large basis set. To ensure that we have a sufficiently
large basis set for this purpose, $220$ basis functions were used for
each normal mode in this work.

\subsection{Thermal expansion}

The high temperatures present in the lower mantle make it important to
consider the effects of thermal expansion. This is most often done
using the quasiharmonic method, where the free energy is calculated as
a function of temperature within the harmonic approximation for
several different values of the lattice parameters. Fitting to these
results then allows the minimum of free energy at a given temperature
to be found as a function of these lattice parameters. It is also
possible to go beyond the quasiharmonic approximation within the VSCF
method used in this work; this involves using the calculated nuclear
wavefunctions to compute the vibrational average of the stress tensor,
adjusting the lattice parameters accordingly, and repeating the
anharmonic calculation until convergence of the lattice parameters is
reached. However, using the VSCF method in this way involves mapping
the BO surface several times at different volumes, making it
significantly more expensive than the quasiharmonic method, which
itself can be expensive. In this work, we instead use previous
experimental data\cite{noguchi_high-temperature_2012} to estimate the
thermal expansion along the geotherm, and consider the results in the
light of the results of the full anharmonic calculations.

\section{Results} \label{sec:Results}

\subsection{Structures investigated}

To obtain a picture of how CaSiO$_3$ behaves across the bottom of the
lower mantle, four points on the geotherm were considered,
corresponding to external pressures of $100$, $128$, $132$, and
$135$~GPa. $100$~GPa was chosen as a representative point in the main
lower mantle, $128$~GPa as corresponding to the top of the D$''$
layer, $132$~GPa as a point in the middle of the D$''$ layer, and
$135$~GPa as corresponding to the CMB. As shown in
Fig.\ \ref{fig:Geotherm}, each of these pressures is associated with a
range of temperatures on the geotherm: $2255$-$2540$, $2450$-$2755$,
$2765$-$3190$, and $3000$-$3520$~K correspond to $100$, $128$, $132$,
and $135$~GPa respectively. In this work, the vibrational free energy
and relative stabilities are calculated at the maximum and minimum
temperature on the geotherm for each pressure, as well as at zero
temperature. Fig.\ \ref{fig:Geotherm} also shows that there are no
discontinuities in the gradient of the geotherm, associated with
changes in phase or composition, in the upper part of the lower
mantle. This implies that the results for $100$~GPa should be
qualitatively applicable to the rest of the lower mantle, but further
first-principles work would be required to obtain a quantitative
description.

To obtain appropriate lattice constants for each pressure, the perfect
cubic perovskite structure was allowed to relax whilst maintaining its
symmetry. This led to zero temperature lattice constants of $3.300$,
$3.251$, $3.245$, and $3.240$~\r{A} for $100$, $128$, $132$, and
$135$~GPa respectively. These compare well with experimental results
for the unit cell volume at high pressures and temperatures, which
imply a cubic lattice constant of $3.239$~\r{A} at $119.1$~GPa and
$704$~K\cite{noguchi_high-temperature_2012}. The slight overestimation
is typical for GGA-based DFT functionals such as
PBE\cite{haas_calculation_2009}.

\begin{figure*}
  \centering
  \subcaptionbox{\emph{Pm$\bar{3}$m}}{
    \includegraphics[trim={6.2cm 10.95cm 2.75cm 6.6cm},clip,width=0.3\textwidth]{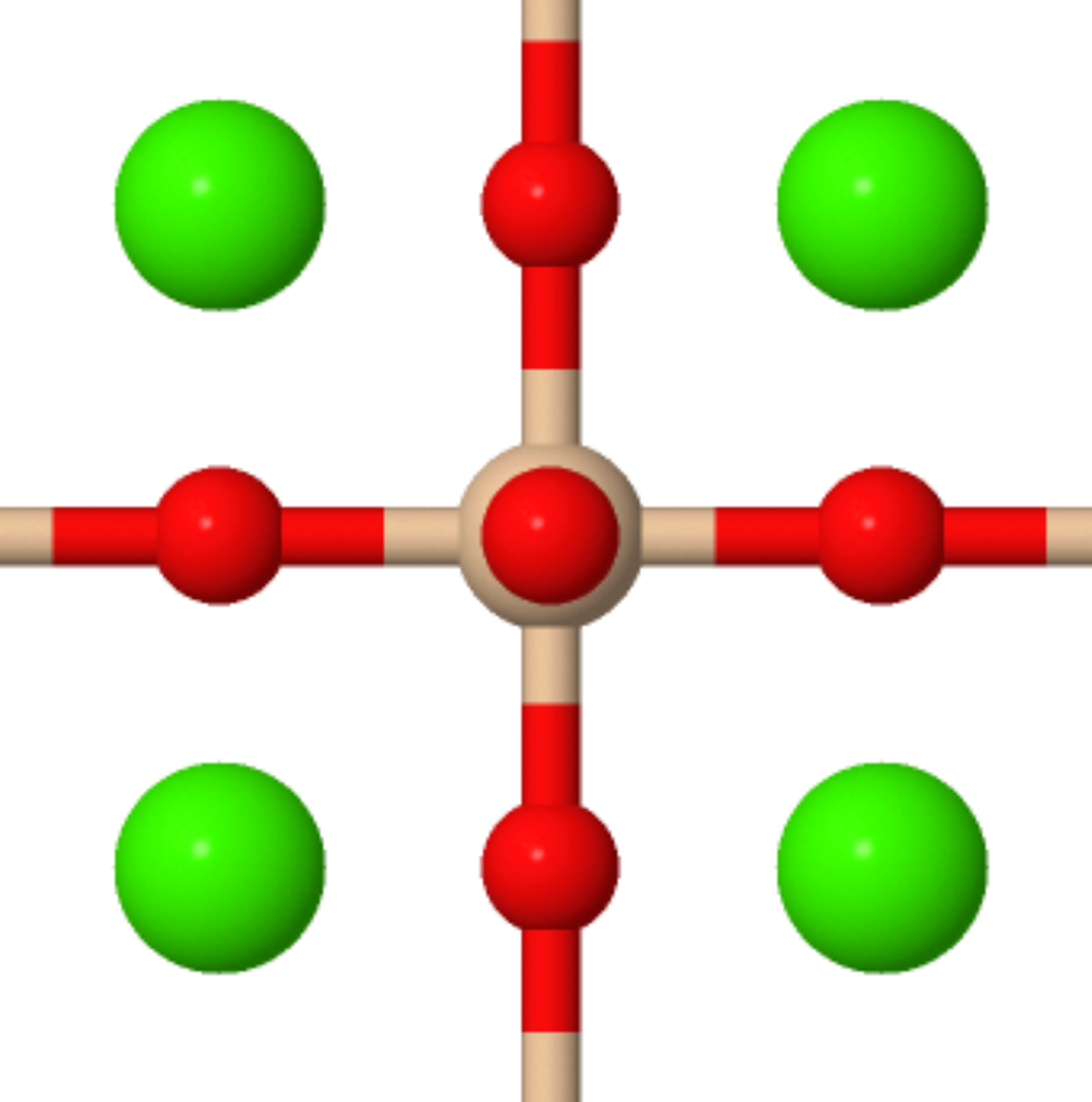}
  }
  ~
  \subcaptionbox{\emph{I4/mcm}}{
    \includegraphics[trim={5.95cm 7.4cm 3cm 10.2cm},clip,width=0.3\textwidth]{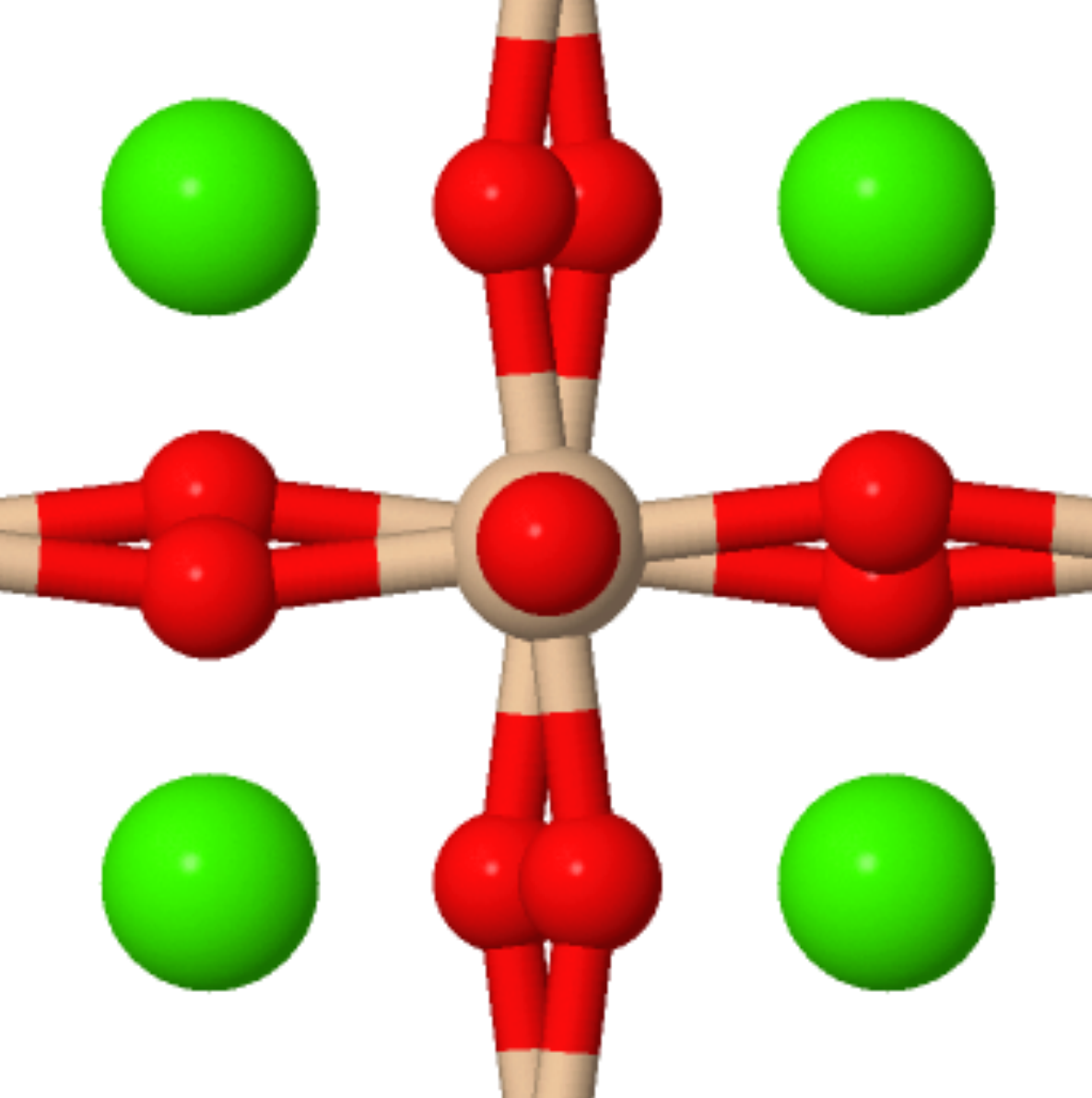}
  }
  ~
  \subcaptionbox{\emph{C2/m}}{
    \includegraphics[trim={5.95cm 11.1cm 3cm 6.6cm},clip,width=0.3\textwidth]{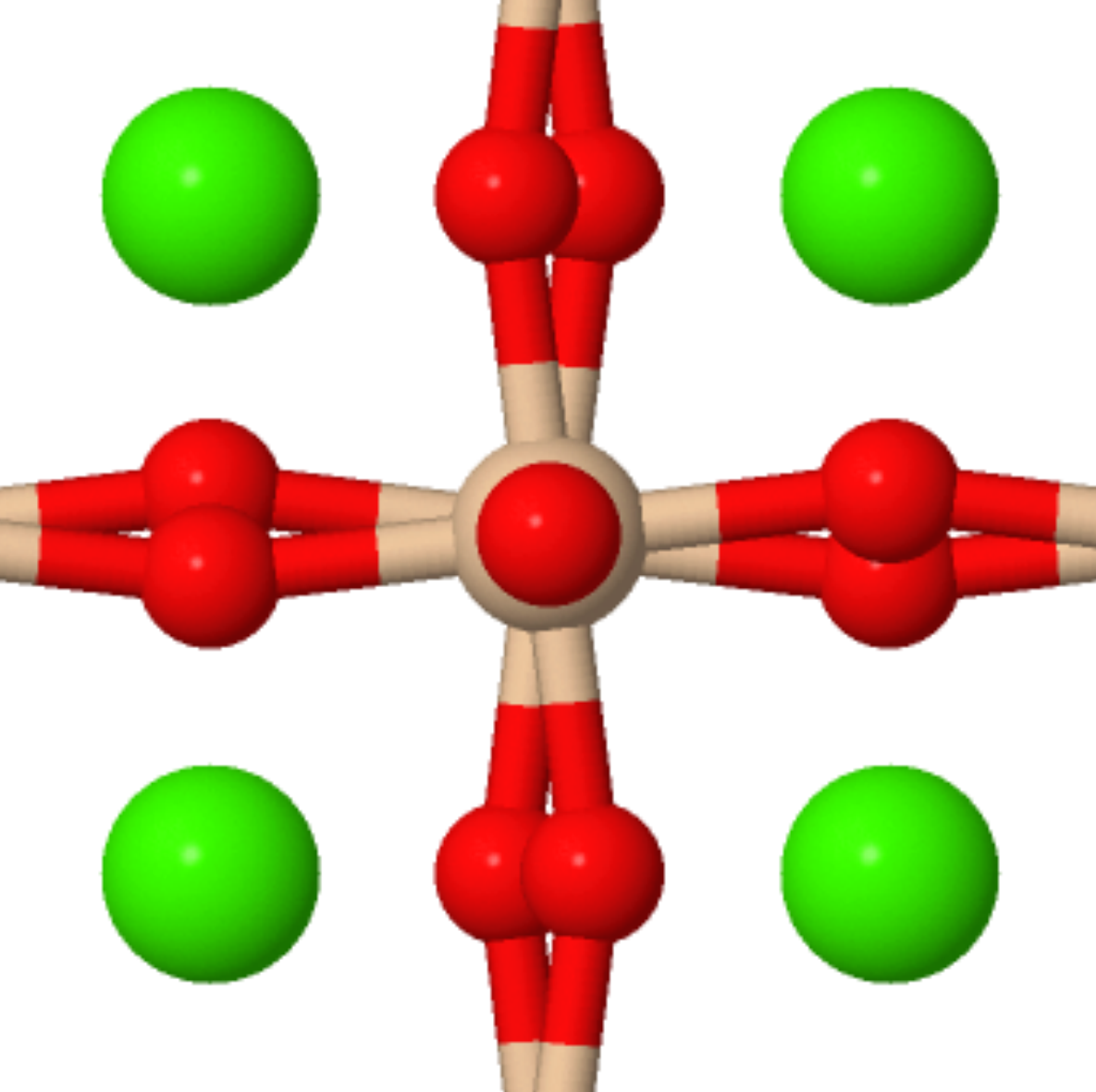}
  }
  \caption{The crystal structures of calcium silicate considered in
    this work. a) shows the basic cubic perovskite \emph{Pm$\bar{3}$m}
    structure, viewed along any of the Cartesian axes. b) shows the
    distorted \emph{I4/mcm} structure, viewed along the $y$-axis (the
    cubic structure's $b$-axis). c) shows the distorted \emph{C2/m}
    structure, viewed along the $z$-axis (the cubic structure's
    $c$-axis). In all figures, Ca is green, O is red and Si is brown.}
  \label{fig:DistortedStructures}
\end{figure*}

Once the relaxed lattice constants had been obtained, the next task
was to find appropriate distorted structures to compare against the
perfect cubic structure. Many different structures have been
considered in previous
work\cite{shim_tetragonal_2002,stixrude_phase_2007,komabayashi_phase_2007,caracas_theoretical_2006,uchida_non-cubic_2009,magyari-kope_low-temperature_2002,adams_ab_2006},
and it is not feasible to do a full anharmonic calculation on every
one of these. Instead, we obtained distorted structures via a more
rigorous method -- by following the soft modes present in the cubic
structure. A harmonic vibrational calculation was conducted for each
of the cubic structures, sampling the vibrational Brillouin zone (BZ)
with a $4\times4\times4$ grid using the non-diagonal supercells
method\cite{lloyd-williams_lattice_2015}. This size of grid was chosen
as previous work has shown that the strongest soft modes lie at the
$M$ and $R$ symmetry points, which have vibrational BZ co-ordinates
$(0, 0.5, 0.5)$ and $(0.5, 0.5, 0.5)$ respectively. Soft modes were
indeed found at the $M$ and $R$ points, as expected, as well as along
the $T$ line between the $M$ and $R$ points at co-ordinates $(0.25,
0.5, 0.5)$ and $(0.75, 0.5, 0.5)$. For each pressure considered, the
cubic structure was distorted to follow each of these soft modes and
the atomic positions and lattice parameters were allowed to relax. In
each case, this resulted in the same set of distorted structures, each
associated with a high symmetry point in the vibrational BZ, which are
summarised in Table \ref{tab:DistortedStructures}. Visualisations of
the structures are also presented in
Fig.\ \ref{fig:DistortedStructures}. At all pressures, the two lowest
enthalpy structures were the \emph{C2/m} and \emph{I4/mcm} structures
originating from the $R$ point. These structures were therefore
selected for a full anharmonic calculation and comparison with the
cubic structure, and will be the only distorted structures discussed
in the following sections.

\begin{table*}[t]
  \centering
  \begin{tabular}{ | c  c | c | c | c | c | c | }
    \cline{4-7}
    \multicolumn{3}{ c }{} & \multicolumn{4}{ | c | }{$\Delta H_\text{elec}$ (meV per f.u.)} \\
    \hline
    \multicolumn{2}{ | c | }{BZ point} & Space group & $100$~GPa & $128$~GPa & $132$~GPa & $135$~GPa \\
    \hline
    $M$ & $(0, 0.5, 0.5)$ & \emph{P4/mbm} & $-0.20$ & $-2.82$ & $-0.21$ & $-0.3$ \\
    $R$ & $(0.5, 0.5, 0.5)$ & \emph{C2/m} & $29.99$ & $35.93$ & $36.90$ & $37.51$ \\
    $R$ & $(0.5, 0.5, 0.5)$ & \emph{I4/mcm} & $29.94$ & $35.91$ & $36.91$ & $37.52$ \\
    $T$ & $(0.75, 0.5, 0.5)$ & \emph{I4cm} & $9.05$ & $10.34$ & $10.62$ & $10.69$ \\
    $T$ & $(0.25, 0.5, 0.5)$ & \emph{I4/mcm} & $9.05$ & $10.34$ & $10.62$ & $10.69$ \\
    \hline
    \multicolumn{7}{ c }{} \\
    \cline{4-7}
    \multicolumn{3}{ c }{} & \multicolumn{4}{ | c | }{$\Delta H_\text{elec}$ (eV per f.u.)} \\
    \hline
    \multicolumn{2}{ | c | }{Post-perovskite} & \emph{Cmcm} & $-0.660$ & $-0.658$ & $-0.656$ & $-0.655$ \\
    \hline
  \end{tabular}
  \caption{Details of the distorted structures found in this work,
    including their space groups, the locations of their associated
    soft modes in the vibrational Brillouin zone, and their enthalpies
    relative to the cubic perovskite structure at each pressure. The
    enthalpies here are purely electronic, with no vibrational
    contribution. $\Delta H_\text{elec}$ represents the difference in
    enthalpy per formula unit between the distorted structure and the
    cubic perovskite structure. A positive value for $\Delta
    H_\text{elec}$ means that the distorted structure is lower in
    enthalpy than the cubic structure, whilst a negative value means it
    is higher in enthalpy.}
  \label{tab:DistortedStructures}
\end{table*}

One addition was made to this rigorous method of obtaining alternative
structures, in order to investigate the stability of the \emph{Cmcm}
post-perovskite structure of MgSiO$_3$ when applied to
CaSiO$_3$. CaSiO$_3$ in this structure was allowed to relax, whilst
maintaining symmetry, at the different pressures selected. The
properties of the resulting structures are also noted in Table
\ref{tab:DistortedStructures} and presented in
Fig.\ \ref{fig:DistortedStructures}. As can be seen, the
post-perovskite structure was found to be more than $0.65$~eV higher
in enthalpy than the cubic perovskite phase across all phases, making
it energetically uncompetitive. From this, we can conclude that
CaSiO$_3$ does not undergo a post-perovskite phase transition like
MgSiO$_3$, and we therefore do not mention this structure in the rest
of this work. The structures used in this work, not including the
post-perovskite structure, can be found in \texttt{.cif} format in the
Supplemental Material\cite{supplementary_material}.

\subsection{Anharmonic calculations}

\subsubsection{Vibrational Brillouin zone sampling}

A key consideration when undertaking vibrational calculations such as
the ones presented here is the sampling of the vibrational Brillouin
zone. The finer this sampling is, the more accurate the calculations
should be, but they will also increase significantly in
expense. Although we use the non-diagonal supercells
method\cite{lloyd-williams_lattice_2015} to reduce this, as mentioned
previously, it is still necessary to find a balance between accuracy
and cost. In this work, we use an $8\times8\times8$ grid sampling of
the vibrational Brillouin zone for the cubic structures, and a
$4\times4\times4$ grid for the distorted structures.

\subsubsection{Mapping amplitudes}

The very high temperatures present in the lower mantle mean that the
nuclei will have enough energy to explore the Born-Oppenheimer surface
out to large amplitudes, and therefore in order to obtain accurate
results our first-principles mapping of the BO surface must do so
too. In order to ensure this, the BO surface was mapped along each
normal mode up to a maximum amplitude of at least $5\sqrt{\langle
  u_i^2 \rangle}$, where $\langle u_i^2 \rangle = \frac{1}{\omega_i}
\left( \frac{1}{e^{\beta\omega_i}-1} + \frac{1}{2} \right)$ is the
harmonic expectation value of the mode amplitude squared. In several
cases this was insufficient to reach energies at least $k_BT$ higher
than the reference state, and in these cases the mapping was extended
until this energy was obtained. This allowed us to ensure that the
most thermally relevant part of the BO surface was mapped from
directly from first-principles.

Although this part of the BO surface is the most important to get
correct, it is not the only part that participates in the
calculations. A key part of the VSCF method is the integral over the
mode amplitudes of the product of the nuclear probability density (the
modulus squared of the wavefunction) and the BO surface. As we are
describing the nuclear wavefunction of each normal mode using $220$
harmonic oscillator basis functions, some of which may extend
significantly beyond the amplitude already mapped explicitly, it is
necessary to extend the limits of the integration when integrating
over these basis functions. This means it is also necessary to have an
expression for the BO surface at these amplitudes. To this end, we
extend the BO surface by assuming it has a quadratic form beyond the
explicitly mapped region, as we observe that the mapped BO surface has
a quadratic form at the extremes of the mapping. A different quadratic
fit is used for extending the BO surface in the negative and positive
directions, and for each fit the parameters are found by fitting to
the three mapped points with the most negative/positive amplitude.

With this extension of the BO surface in hand, the limits of
integration for an integral over a particular harmonic oscillator
basis function, quantum number $n$ and frequency $\omega$, are taken
to be either
\begin{itemize}
\item the amplitudes that have already been explicitly mapped, or
\item the maximum amplitude of a classical harmonic oscillator with
  the same energy as the quantum harmonic oscillator state associated
  with the basis function, given by $\sqrt{\frac{2n+1}{\omega}}$,
\end{itemize}
whichever is greater. This is because the correspondence principle
tells us that as $n$ increases, the quantum harmonic oscillator states
describe probability distributions that look more and more like the
classical trajectory, which is bounded at the amplitude given above.

\subsection{Relative stabilities}

\begin{table*}[t]
  \centering
  \begin{tabular}{ c | c | c c c | c c c | c c c | c c c | }
    \multicolumn{1}{ c }{} & \multicolumn{1}{ c }{} & \multicolumn{12}{ c }{$\Delta G$ (eV per f.u.)} \\
    \cline{3-14}
    \multicolumn{1}{ c }{} & \multicolumn{1}{ c | }{Pressure (GPa)} & \multicolumn{3}{ | c | }{$100$} & \multicolumn{3}{ | c | }{$128$} & \multicolumn{3}{ | c | }{$132$} & \multicolumn{3}{ | c | }{$135$} \\
    \cline{3-14}
    \multicolumn{1}{ c }{} & \multicolumn{1}{ c | }{Temperature (K)} & $0$ & $2255$ & $2540$ & $0$ & $2450$ & $2755$ & $0$ & $2765$ & $3190$ & $0$ & $3000$ & $3520$ \\
    \cline{2-14}
    \multirow{2}{*}{Structure} & \emph{I4/mcm} & $0.007$ & $-0.080$ & $-0.093$ & $0.012$ & $-0.083$ & $-0.096$ & $0.020$ & $-0.090$ & $-0.108$ & $-0.003$ & $-0.101$ & $-0.119$ \\
    & \emph{C2/m} & $0.007$ & $-0.075$ & $-0.087$ & $0.012$ & $-0.086$ & $-0.100$ & $0.021$ & $-0.089$ & $-0.108$ & $-0.006$ & $-0.122$ & $-0.144$ \\
    \cline{2-14}
  \end{tabular}
  \caption{The relative stabilities of the distorted structures
    relative to the cubic perovskite structure at each pressure, at
    zero temperature and at the upper and lower bounds of the
    temperature range associated with that pressure on the
    geotherm. $\Delta G$ represents the difference in Gibbs free
    energy per formula unit between the distorted structure and the
    cubic perovskite structure. A positive value for $\Delta G$ means
    that the distorted structure is lower in free energy than the
    cubic structure, whilst a negative value means it is higher in
    free energy.}
  \label{tab:RelStabFiniteTemp}
\end{table*}

\begin{figure}
\centering
\includegraphics[width=0.45\textwidth]{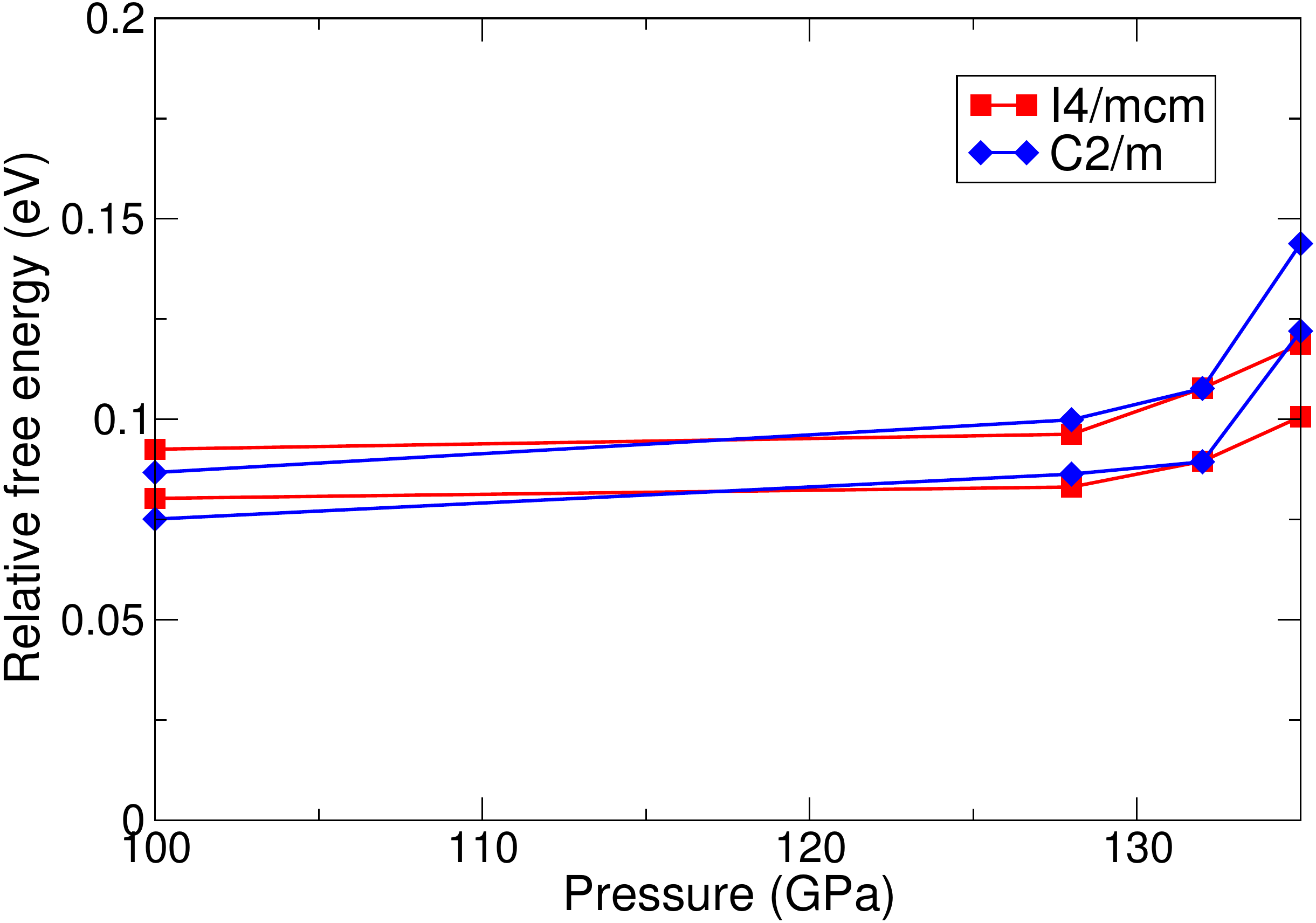}
\caption{The main data from Table \ref{tab:RelStabFiniteTemp}
  presented visually -- the Gibbs free energy per formula unit of the
  distorted structures relative to the cubic perovskite structure at
  each pressure, and at the upper and lower bounds of the temperature
  range associated with that pressure on the geotherm.}
\label{fig:RelStabFiniteTemp}
\end{figure}

Table \ref{tab:RelStabFiniteTemp} presents the main results of this
work -- that is, the relative stabilities of the structures considered
at several different points on the geotherm. Figure
\ref{fig:RelStabFiniteTemp} presents the same information graphically.

The key result here is that at the temperatures and pressures found in
the lower part of the lower mantle, the cubic structure of CaSiO$_3$
is always more stable, and in fact becomes more stable the deeper into
the mantle one goes, from $0.075$~meV -- at $100$~GPa -- to
$0.122$~meV -- at $135$~GPa -- lower in free energy than the lowest
energy distorted structure (at the lowest temperature considered for
each pressure). This supports previous work that suggests that calcium
silicate is cubic in the lower
mantle\cite{sun_dynamic_2014,komabayashi_phase_2007,stixrude_phase_2007,adams_ab_2006,noguchi_high-temperature_2012},
utilising a novel high accuracy first-principles method to complement
previous work. The distortion that is lowest in free energy changes as
the pressure and temperature increase, going from \emph{C2/m} at
$100$~GPa to \emph{I4/mcm} at an estimated crossover pressure of
$117$~GPa.

At zero temperature, the zero point vibrational energy has the effect
of closing the gap between the cubic structure and the distorted
structures, which would be significantly lower in free energy if this
contribution were not included. At $135$~GPa, this effect is actually
strong enough to alter the relative free energies by more than
$0.6$~eV and make the distorted structures higher in free energy than
the cubic structure, making the cubic structure thermodynamically
stable (although still dynamically unstable due to the presence of
soft modes).

\subsection{Mapped phonon modes}

\begin{figure}
  \centering
  \subcaptionbox{\label{fig:CubicSoftMode}}{
    \includegraphics[width=0.45\textwidth]{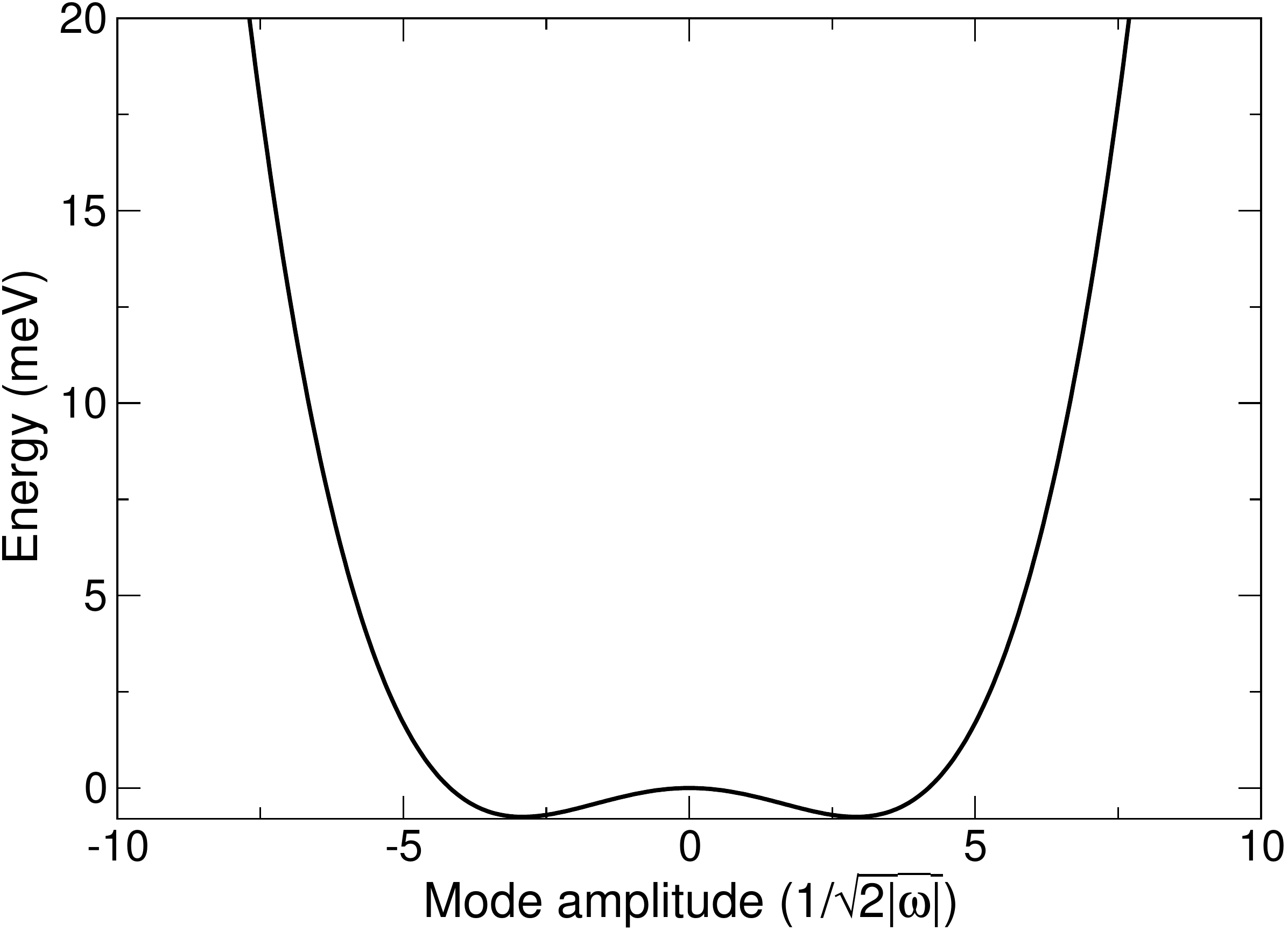}
  }
  ~
  \subcaptionbox{\label{fig:DistortedDoubleWell}}{
    \includegraphics[width=0.45\textwidth]{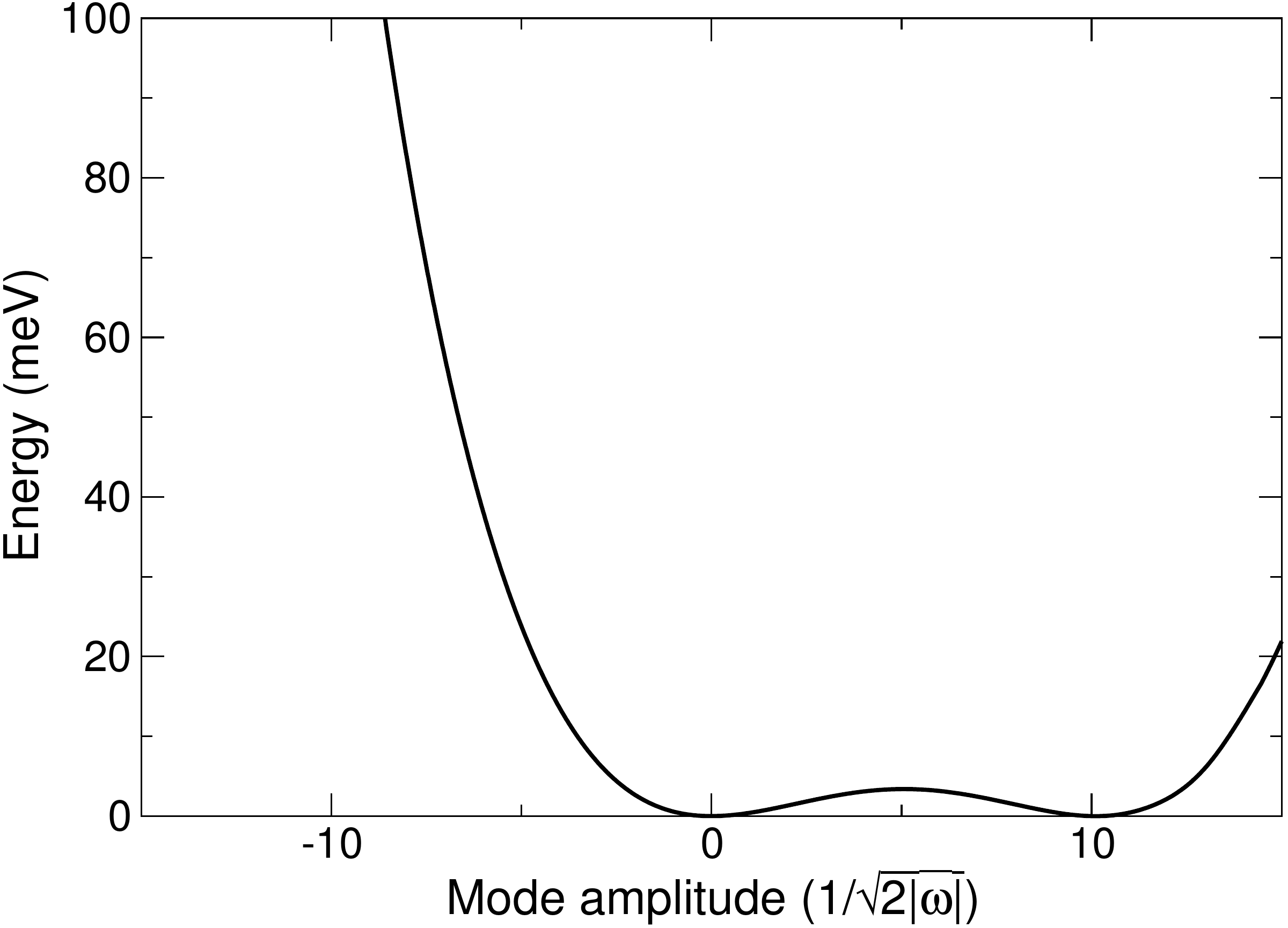}
  }
  \caption{The same soft mode mapped twice, using different reference
    structures. a) shows the soft mode that was followed to obtain the
    \emph{I4/mcm} structure at $100$~GPa, with the cubic structure at
    $0$ amplitude. b) shows the same mode but mapped with the
    \emph{I4/mcm} structure as the reference structure at $0$
    amplitude. In both cases, the amplitude is measured in units of
    $\frac{1}{\sqrt{2|\omega|}}$, where $\omega$ is the frequency of
    the soft mode in the cubic structure.}
  \label{fig:SoftMode}
\end{figure}

Figure \ref{fig:SoftMode} shows two examples of phonon modes mapped as
part of these calculations. Figure \ref{fig:CubicSoftMode} shows one
of the soft modes present in the cubic structure at $100$~GPa, whilst
Fig.\ \ref{fig:DistortedDoubleWell} shows the same mode, but mapped
with the \emph{I4/mcm} distorted structure as the reference structure
at $100$~GPa. Both show the characteristic double well potential of a
soft mode. Both wells in Fig.\ \ref{fig:DistortedDoubleWell}
correspond to distortions with the same symmetry, whilst the structure
corresponding to the maximum is cubic. This presents the possibility
of the system being able to `hop' between these wells if it has enough
energy, resulting in the structure becoming cubic on average, although
instantaneously the structure is distorted, in a similar way to the
dynamic Jahn-Teller effect\cite{prentice_first-principles_2017}. At
the very high temperatures seen in the lower mantle, this is unlikely
to be the root cause of the stability of the cubic structure, as the
depth of these wells will become negligible compared to the thermal
energy, but at intermediate temperatures this may be an important
effect in stabilising the cubic structure.

\subsection{Thermal expansion}

\begin{table*}[t]
  \centering
  \begin{tabular}{ c | c c c | c c c | c c c | c c c | }
    \cline{2-13}
    Pressure (GPa) & \multicolumn{3}{ | c | }{$100$} & \multicolumn{3}{ | c | }{$128$} & \multicolumn{3}{ | c | }{$132$} & \multicolumn{3}{ | c | }{$135$} \\
    \cline{2-13}
    Temperature (K) & $0$ & $2255$ & $2540$ & $0$ & $2450$ & $2755$ & $0$ & $2765$ & $3190$ & $0$ & $3000$ & $3520$ \\
    \cline{2-13}
    $a(P,T)$ (\r{A}) & $3.300$ & $3.342$ & $3.348$ & $3.251$ & $3.288$ & $3.293$ & $3.245$ & $3.286$ & $3.293$ & $3.240$ & $3.283$ & $3.292$ \\
    \cline{2-13}
    Increase (\%) & -- & $1.27$ & $1.45$ & -- & $1.12$ & $1.28$ & -- & $1.25$ & $1.47$ & -- & $1.34$ & $1.61$ \\
    \cline{2-13}
  \end{tabular}
  \caption{The DFT lattice parameters of the cubic perovskite
    structure and the thermally expanded lattice parameters estimated
    from these using experimental
    data\cite{noguchi_high-temperature_2012}. $a(P,0)$ is the lattice
    parameter at a given pressure $P$ and temperature $T$. $a(P,0)$ is
    the DFT lattice parameter previously calculated for that structure
    at $0$~K. All other lattice parameters are estimated using the
    results of Ref.\ \onlinecite{noguchi_high-temperature_2012}. The
    increase row shows the estimated percentage increase in the
    lattice parameter compared to the $0$~K value for that pressure.}
  \label{tab:ExpThermalExpansion}
\end{table*}

Previous experimental work by Noguchi \textit{et
  al.}\cite{noguchi_high-temperature_2012} has provided several
different models for a thermal equation of state for cubic calcium
silicate perovskite under lower mantle conditions. Using these models
allows us to link pressure, volume and temperature, and find an
estimate for the volume at a given pressure and temperature, including
the effects of thermal expansion. Here, we use the thermodynamic
thermal pressure model (model 4 in
Ref.\ \onlinecite{noguchi_high-temperature_2012}), which gives the
equation of state as
\begin{align}
  P = &\frac{3}{2} K_0 \left[ \left(\frac{V_0}{V}\right)^{\frac{7}{3}} - \left(\frac{V_0}{V}\right)^{\frac{5}{3}} \right] + \alpha_0 K_0 (T - T_0) \\ 
  &+ K_1 (T- T_0) \ln \left(\frac{V_0}{V}\right) ~, \nonumber
\end{align}
where $K_0=207\pm4$~GPa, $V_0=46.5\pm0.1$~\r{A}$^3$,
$\alpha_0=5.7\times10^{-5}$~K$^{-1}$,
$K_1=-0.010\pm0.004$~GPa\,K$^{-1}$, and $T_0=700$~K. $700$~K is used
as the reference temperature, as they found this was above the
transition temperature to the cubic phase. $V_0$ is the reference
volume at this temperature and $1$~bar$=100$~MPa, but to account for
the error in the volume calculated by DFT compared to the experimental
volume, we solve this equation for $\nu=\frac{V_0}{V}$ instead of $V$.

To estimate the thermal expansion, we solved this equation for each
pressure at the high and low temperatures on the geotherm, as well as
at $0$~K. Although the model uses a reference temperature of $700$~K,
we found that the variation in the DFT volumes at $0$~K with pressure
matched the prediction of the model very well, giving us confidence
that the model was still applicable at $0$~K (at least theoretically,
as the cubic phase is unstable at zero temperature). We can then use
these results to estimate the effect of thermal expansion on the DFT
lattice constant $a$ for each set of conditions, using the formula
\begin{equation}
  a(P,T) = \sqrt[3]{\frac{\nu(P,0)}{\nu(P,T)}} a(P,0) ~.
\end{equation}

The results of this are presented in Table
\ref{tab:ExpThermalExpansion}. It can be seen that there is an
expansion of between 1 and 2\% in all cases. However, with the
exception of the data for $100$~GPa, all the lattice constants remain
within the range of lattice constants already considered in this
work. As Fig.\ \ref{fig:RelStabFiniteTemp} and Table
\ref{tab:RelStabFiniteTemp} show, the cubic phase is stable across the
whole range, meaning that the main conclusion of this work is still
valid -- CaSiO$_3$ takes up the cubic structure throughout
the lower region of the lower mantle. To quantify
the stability of the cubic structure more exactly, a harmonic or full
anharmonic calculation could be performed at these expanded lattice
parameters, and a correction made to the non-thermally expanded
result.

\subsection{Effect of impurities}

The calculations presented here focuses on pure calcium silicate,
neglecting the effect of impurities. This is the assumption made by
most previous theoretical work, and is justified by experimental
results on both naturally occurring calcium
silicate\cite{nestola_casio3_2018} and laboratory
samples\cite{ono_phase_2004,kesson_mineralogy_1998,irifune_absence_1994}. These
experimental results imply that naturally occurring calcium silicate
perovskite is very pure ($>$90\%), as impurities tend to reside in the
magnesium silicate perovskite that exists alongside CaSiO$_3$ in the
lower mantle. At the very highest pressure present in the lower
mantle, however, previous work has suggested that there is a limited
amount of magnesium silicate perovskite in solid solution within
calcium silicate (up to 20\% per
mole)\cite{kesson_mineralogy_1998}. Although this could potentially
cause the structure of CaSiO$_3$ to distort, comparisons to similar
mixtures of perovskites suggest it is more likely that this limited
amount of mixing would not be quite enough to result in a change of
structure\cite{ceh_phase_1987}, resulting in the structure of
CaSiO$_3$ remaining cubic.

\section{Conclusions} \label{sec:Summary}

In summary, we have presented, to the best of our knowledge, the most
in-depth first-principles study of calcium silicate perovskite and its
structure yet. We have thoroughly mapped the Born-Oppenheimer surface
of this key mantle material in both the high symmetry cubic phase and
in two competing distorted phases, and used this data to conduct high
accuracy anharmonic vibrational calculations of the free energy. These
calculations show that, even down in the very depths of the mantle,
calcium silicate takes up the high symmetry cubic structure. This
supports previous work that suggests the structure is
cubic\cite{sun_dynamic_2014,komabayashi_phase_2007,stixrude_phase_2007,adams_ab_2006,noguchi_high-temperature_2012},
and further demonstrates the validity of the first-principles VSCF
method. Although our calculations do not explicitly include any
thermal expansion, we suggest that this result is robust against the
effect of thermal expansion, which was calculated using previous
experimental data. Future work on this area could include the effect
of impurities or thermal expansion fully from first principles, using
either a quasiharmonic or fully anharmonic method. Anharmonic
first-principles calculations could also be applied to other materials
in the interior of Earth, and potentially other planets as well. This
would both help improve the understanding of geophysical phenomena,
and inform future experiments on these materials.

\section{Acknowledgements}

J.C.A.P.  and R.J.N. are grateful to the Engineering and Physical
Sciences Research Council (EPSRC) of the UK for financial support
[EP/P034616/1]. Computational resources were provided by:
\begin{itemize}
\item the Cambridge Service for Data Driven Discovery (CSD3) operated
  by the University of Cambridge Research Computing Service, provided
  by Dell EMC and Intel using Tier-2 funding from the Engineering and
  Physical Sciences Research Council (capital grant EP/P020259/1), and
  DiRAC funding from the Science and Technology Facilities Council,
\item the ARCHER UK National Supercomputing Service, for which access
  was obtained via the UKCP consortium [EP/K014560/1],
\item the Research Center for Advanced Computing Infrastructure at
  JAIST,
\item and the UK Materials and Molecular Modelling Hub, which is
  partially funded by EPSRC (EP/P020194/1), for which access was
  obtained via the UKCP consortium and funded by EPSRC grant ref
  EP/P022561/1.
\end{itemize}

\bibliography{CaSiO3Bib}

\end{document}


\title{Supplementary Information: First principles anharmonic vibrational study of the structure of calcium silicate perovskite under lower mantle conditions}

\author{Joseph C.\ A.\ Prentice \and Ryo Maezono \and R.\ J.\ Needs}

\date{\today}

\maketitle

\section{Pseudopotentials}

All density functional theory calculations were performed using CASTEP
version 16.1, and its own "on-the-fly" ultrasoft pseudopotentials. The
definition strings for the pseudopotentials were:
\begin{itemize}
\item O: \\
  \texttt{2|1.1|15|18|20|20:21(qc=7)}
\item Si: \\
  \texttt{3|1.8|5|6|7|30:31:32}
\item Ca: \\
  \texttt{3|2.0|7|8|9|30U:40:31:32}
\end{itemize}

\section{Equilibrium unit cell configurations}

The unit cells for the structures used in this work, containing the
atoms at their equilibrium positions for each pressure considered, are
provided in \texttt{.cif} format, with the files labelled as
\texttt{<Space group>\_<Pressure>.cif}, e.g.\ \texttt{Pm-3m\_100.cif}.

\section{The VSCF method}

The vibrational self-consistent field (VSCF) method used in this work
was proposed in the context of solid state systems in Monserrat et
al., Phys.\ Rev.\ B \textbf{87}, 144302 (2013). Here, we go into more
detail regarding the method and how it is practically used in this
work.

The main approximation underlying the VSCF method we use is the
principal axes approximation (PAA). This assumes that the harmonic
phonon normal co-ordinates of the system constitute a reasonable basis
to expand the Born-Oppenheimer surface in. We then perform this
expansion of the BO surface, with successive terms having higher and
higher dimensionality in these co-ordinates:
\begin{equation}
  E_{\text{PAA}} (\mathbf{u}) = E_{\text{BO}}(\mathbf{0}) + \sum_{\nu\mathbf{q}}
  V_{\nu\mathbf{q}} \left(u_{\nu\mathbf{q}}\right) + \frac{1}{2}
  \sum_{\nu\mathbf{q}} \sum_{\nu'\mathbf{q}'\neq \nu\mathbf{q}}
  V_{\nu\mathbf{q};\nu'\mathbf{q}'} \left(u_{\nu\mathbf{q}},
  u_{\nu'\mathbf{q}'}\right) ~.
\end{equation}
$E_{\text{PAA}}(\mathbf{u})$ is the PAA to the full BO surface
$E_{\text{BO}} (\mathbf{u})$. $\mathbf{u}$ is a collective variable of
the harmonic normal mode amplitudes $u_{\nu\mathbf{q}}$, which are
labelled according to phonon branch $\nu$ and wavevector $q$. The 1-D
terms (called $V_1$ for short in the main text) are defined as
\begin{equation}
  V_{\nu\mathbf{q}} \left(u_{\nu\mathbf{q}}\right) = E_{\text{BO}}
  \left(0,\dots,0,u_{\nu\mathbf{q}},0,\dots,0\right) -
  E_{\text{BO}}(\mathbf{0}) ~,
\end{equation}
the 2-D terms (called $V_2$ in the main text) are defined as
\begin{align}
  V_{\nu\mathbf{q};\nu'\mathbf{q}'}
  \left(u_{\nu\mathbf{q}},u_{\nu'\mathbf{q}'}\right) =& E_{\text{BO}}
  \left(0,\dots,0,u_{\nu\mathbf{q}},0,\dots,0,u_{\nu'\mathbf{q}'},0,\dots,0\right) \\
  &- V_{\nu\mathbf{q}} - V_{\nu'\mathbf{q}'} - E_{\text{BO}}(\mathbf{0}) ~, \nonumber
\end{align}
and so on. Anharmonicity is already included within the 1-D term, as
it is not constrained to be quadratic in form. We typically truncate
this expansion, and in this work we only include the 1-D terms in our
mapping of the BO surface. The terms of this expansion are then mapped
by freezing in one (or more) phonons at a series of amplitudes and
calculating the electronic energy using DFT, mapping a series of
points on the BO surface. Splines are then used to fit to these
points, giving a first principles mapping of the BO surface that can
take any functional form.

Given the BO surface from the PAA, we can now solve the vibrational
Schr\"odinger equation. We assume the overall vibrational wavefunction
is a Hartree product of single-mode wavefunctions $\phi_{\nu\mathbf{q}}
\left(u_{\nu\mathbf{q}}\right)$, but these single-mode wavefunctions are
not just simple harmonic oscillator wavefunctions, as would be the
case in the harmonic approximation. We then solve the vibrational
Schr\"odinger equation using a mean field, or self-consistent,
approach, hence the name of the method. This leads to the VSCF
equations
\begin{equation}
  \left( -\frac{1}{2} \frac{\partial^2}{\partial u_{\nu\mathbf{q}}^2} +
  \bar{V}_{\nu\mathbf{q}} \left(u_{\nu\mathbf{q}}\right) \right)
  \phi_{\nu\mathbf{q}} \left(u_{\nu\mathbf{q}}\right) =
  \lambda_{\nu\mathbf{q}} \phi_{\nu\mathbf{q}}
  \left(u_{\nu\mathbf{q}}\right) ~,
\end{equation}
where the mean-field potential is defined as
\begin{equation}
  \bar{V}_{\nu\mathbf{q}} \left(u_{\nu\mathbf{q}}\right) = \int
  \prod_{\nu'\mathbf{q}'\neq \nu\mathbf{q}} \text{d}u_{\nu\mathbf{q}} \left|
  \phi_{\nu\mathbf{q}} \left(u_{\nu\mathbf{q}}\right) \right|^2
  E_{\text{PAA}} (\mathbf{u}) ~.
\end{equation}
Solving these equations self-consistently produces a set of
single-mode states, including both ground and excited states, for each
mode, which can each be labelled by the index $s_{\nu\mathbf{q}}$
(collective variable $\mathbf{s}$). The vibrational wavefunction of an
overall state $\mathbf{s}$ is then
\begin{equation}
  \Phi_{\mathbf{s}} (\mathbf{u}) = \prod_{\nu\mathbf{q}}
  \phi_{\nu\mathbf{q};s_{\nu\mathbf{q}}} \left(u_{\nu\mathbf{q}}\right)
\end{equation}
and has energy
\begin{align}
  E_{\mathbf{s}}^\text{(vib)} =&
  \sum_{\nu\mathbf{q}}\lambda_{\nu\mathbf{q};s_{\nu\mathbf{q}}} \\ &+ \int
  \prod_{\nu\mathbf{q}} \text{d}u_{\nu\mathbf{q}} \left|
  \phi_{\nu\mathbf{q};s_{\nu\mathbf{q}}} \left(u_{\nu\mathbf{q}}\right)
  \right|^2 \left( E_{\text{PAA}} (\mathbf{u}) - \sum_{\nu\mathbf{q}}
  \bar{V}_{\nu\mathbf{q}} \left(u_{\nu\mathbf{q}}\right) \right) ~. \nonumber
\end{align}
A perturbation theory at second order can be constructed on the VSCF
equations, using $\left( E_{\text{PAA}} (\mathbf{u}) -
\sum_{\nu\mathbf{q}} \bar{V}_{\nu\mathbf{q}} \left(u_{\nu\mathbf{q}}\right)
\right)$ as the small perturbation to the mean-field Hamiltonian --
the first-order energy correction is included in the expression for
$E_{\mathbf{s}}^\text{(vib)}$ above. In the current work, as we do not
include any 2-D terms in the PAA expansion of the BO surface, this
perturbation is exactly 0, so no perturbation theory is constructed.

We therefore have a set of anharmonic energies and wavefunctions for
all the vibrational states of the system. This means that we can
calculate a partition function
\begin{equation}
  \mathcal{Z} = \sum_{\mathbf{s}} e^{-\beta
    E_{\mathbf{s}}^{\text{(vib)}}} ~, 
\end{equation}
from which we can calculate the Helmholtz free energy
\begin{equation}
  F = -\frac{1}{\beta} \ln \mathcal{Z} ~,
\end{equation}
from which we can calculate the Gibbs free energy by including the
effect of pressure. In a system with no band gap, electronic excited
states would also become important, changing the shape of the BO
surface with temperature, but in a system with a significant band gap
like CaSiO$_3$, the BO surface at zero temperature remains a good
approximation throughout the temperature range considered here.

In practice, we solve the VSCF equations for each mode by writing the
$\phi_{\nu\mathbf{q}} \left(u_{\nu\mathbf{q}}\right)$ as a linear
combination of simple harmonic oscillator wavefunctions (with a
frequency found by fitting a quadratic potential to the mapped BO
surface along the mode concerned). Doing this has several advantages:
these functions are simple to calculate; not too many should be
required for an accurate representation of the anharmonic states, as
they should be reasonable first approximations to the anharmonic
states; and if the Hamiltonian is split into a harmonic part
(including the kinetic part) and an anharmonic correction to the
potential, the bulk of the Hamiltonian matrix can be calculated
analytically. Once the Hamiltonian has been written in this basis, it
is explicitly diagonalised to obtain the anharmonic wavefunctions and
energies.